\documentclass[sigconf,nonacm]{acmart}
\usepackage{multirow}
\usepackage{xspace}
\makeatletter
\g@addto@macro\UrlBreaks{\do\/\do-\do\.\do\_\do\?\do\&\do\=\do\#}
\makeatother
\usepackage{framed}
\usepackage{algorithm}
\usepackage{algorithmic}

\copyrightyear{2026}
\acmYear{2026}
\setcopyright{cc}
\setcctype{by}
\acmConference[CIKM '26]{Proceedings of the 35th ACM International Conference on Information and Knowledge Management}{November 07--11, 2026}{Rome, Italy}
\acmBooktitle{Proceedings of the 35th ACM International Conference on Information and Knowledge Management (CIKM '26), November 07--11, 2026, Rome, Italy}
\acmISBN{979-8-4007-2539-5/2026/11}
\acmDOI{10.1145/3799682.3841109}
\definecolor{promptbg}{gray}{0.94}
\newenvironment{promptbox}%
  {\setlength{\fboxsep}{4pt}%
   \MakeFramed{\advance\hsize-\width\FrameRestore}}%
  {\endMakeFramed}

\newcommand{\sysname}{NASA-EO-Bench\xspace}
\newcommand{\nnname}{NN-SSC\xspace}   
\newcommand{\RK}[1]{\text{R@}#1}
\newcommand{\PK}[1]{\text{P@}#1}
\newcommand{\MAP}{\text{MAP}\xspace}
\newcommand{\MRR}{\text{MRR}\xspace}

\begin{document}

\title{Bringing Agentic Search to Earth Observation Data Discovery}
\renewcommand{\shorttitle}{Agentic Search for Earth Observation Data Discovery}

\author{Minghan Yu}
\authornote{Minghan Yu and Youran Sun contributed equally to this work.}
\orcid{0009-0006-1029-0461}
\affiliation{%
  \department{Department of Mathematics}
  \institution{University of Maryland, College Park}
  \city{College Park}
  \state{MD}
  \country{USA}}
\email{my6489@umd.edu}

\author{Youran Sun}
\orcid{0000-0001-7261-6283}
\authornotemark[1]
\affiliation{%
  \department{Department of Mathematics}
  \institution{University of Maryland, College Park}
  \city{College Park}
  \state{MD}
  \country{USA}}
\email{sun1245@umd.edu}

\author{Chugang Yi}
\orcid{0009-0005-7081-2674}
\affiliation{%
  \department{Department of Mathematics}
  \institution{University of Maryland, College Park}
  \city{College Park}
  \state{MD}
  \country{USA}}
\email{chugang@umd.edu}

\author{Yixin Wen}
\orcid{0000-0003-1618-1208}
\authornote{Yixin Wen and Haizhao Yang are corresponding authors.}
\affiliation{%
  \department{School of Marine and Atmospheric Sciences}
  \institution{Stony Brook University}
  \city{Stony Brook}
  \state{NY}
  \country{USA}}
\email{yixin.wen@stonybrook.edu}

\author{Haizhao Yang}
\authornotemark[2]
\orcid{0000-0002-8408-1754}
\affiliation{%
  \department{Department of Mathematics}
  \institution{University of Maryland, College Park}
  \city{College Park}
  \state{MD}
  \country{USA}}
\affiliation{%
  \department{Department of Computer Science}
  \institution{University of Maryland, College Park}
  \city{College Park}
  \state{MD}
  \country{USA}}
\email{hzyang@umd.edu}
\renewcommand{\shortauthors}{Minghan Yu, Youran Sun, Chugang Yi, Yixin Wen, and Haizhao Yang}

\begin{abstract}
NASA and its data centers hold thousands of geoscience datasets and tools like Worldview, Giovanni, the Science Discovery Engine, and Harmony.
Finding the right one is hard even for domain experts.
We present an agentic search framework for geoscience data discovery that takes a natural-language research query and returns matching datasets and tools.
We demonstrate that, in the era of large language models, the latent value of knowledge graphs (KGs) can be substantially amplified through agentic search.
From the NASA Earth Observation Knowledge Graph (NASA EO-KG) we derive \emph{\sysname}, an open benchmark of 47k query--dataset pairs (21k task-based queries).
A neural scorer fine-tuned on \sysname beats cosine and BM25 baselines.
Further combining it with BM25 via score fusion raises both Recall@10 ($\RK{10}$) and \MRR to over $5\times$ the unadapted cosine baseline.
On top of this supervised pipeline, a zero-shot reranking stage lifts \MRR by 16\%, significant under a paired bootstrap, with no additional training, and autonomous web and arXiv tool use adds a further gain, showing that LLM reasoning is complementary to supervised retrieval.
\end{abstract}



\maketitle
\section{Introduction}
\label{sec:intro}

Earth-observation data discovery is less a problem of data scarcity than of navigating fragmented metadata, tools, and access pathways.
NASA and its affiliated data centers host thousands of datasets across dozens of Distributed Active Archive Centers (DAACs), together with tools such as Worldview, Giovanni, Science Discovery Engine, and Harmony.\footnote{%
  Worldview \url{https://worldview.earthdata.nasa.gov/},
  Giovanni \url{https://giovanni.gsfc.nasa.gov/giovanni/},
  Science Discovery Engine \url{https://science.data.nasa.gov/science-discovery-engine/},
  Harmony \url{https://harmony.earthdata.nasa.gov/}.%
}
This fragmentation makes it hard even for domain experts to locate the data that best matches their own research question.

\begin{figure*}[t!]
  \centering
  \includegraphics[width=\linewidth]{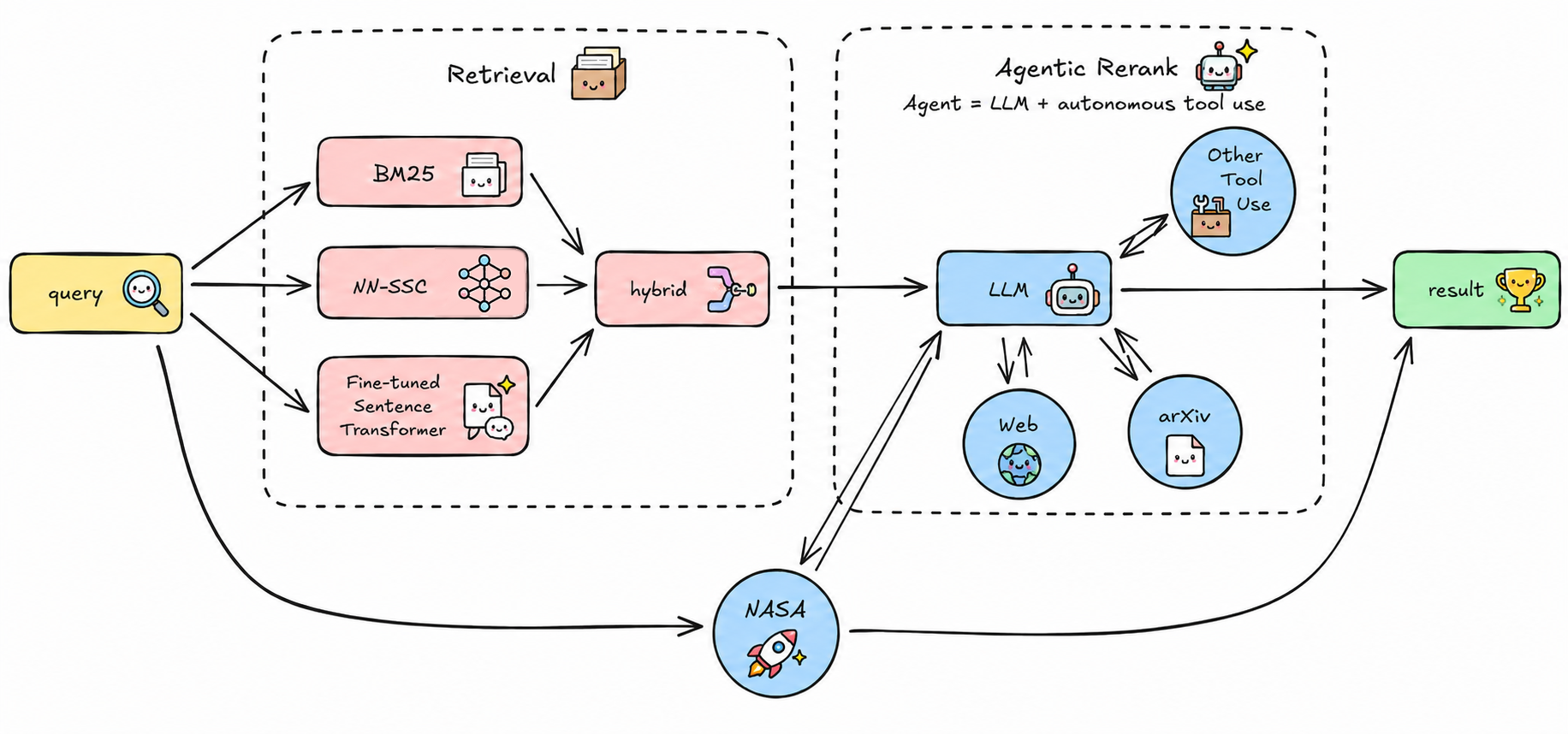}
  \caption{Overview of the three-stage agentic search pipeline (Section~\ref{sec:agent-pipeline}).
  \textbf{Stage 1}: the Router first attempts to resolve each query via NASA official tools (Harmony, the Science Discovery Engine (SDE), WorldView, Giovanni); queries that can be fully answered here terminate early.
  \textbf{Stage 2}: if no official tool suffices, the hybrid BM25 + \nnname retriever surfaces the most relevant datasets from the NASA Common Metadata Repository (CMR) corpus.
  \textbf{Stage 3}: retrieved candidates are reranked by an LLM that autonomously invokes web search and arXiv paper lookup to ground its ranking decisions in external context; the final ranked list is returned to the user.}
  \Description{A left-to-right flow diagram. A yellow query node on the left fans out to three pink retrieval nodes, BM25, NN-SSC, and a fine-tuned sentence transformer, enclosed in a dashed box labelled Retrieval; their arrows converge on a pink hybrid node. A second dashed box labelled Agentic Rerank holds a blue LLM node joined by double-headed arrows to three blue circles for web search, arXiv lookup, and other tool use. A separate blue NASA node sits below both dashed boxes, reached directly from the query and linked back to the LLM, forming a bypass path. Both the LLM and the NASA node feed a green result node on the right.}
  \label{fig:pipeline}
\end{figure*}

A large language model (LLM) offers researchers a natural interface for expressing dataset-retrieval intent in natural language. In a highly domain-specific setting such as geoscience dataset retrieval, however, using an LLM directly for queries faces two cumulative challenges.
First, the pre-training corpora of general LLMs are dominated by general web text and lack the observational data and domain knowledge of geoscience, so domain queries are often neither accurately understood nor directly answerable in a trustworthy way.
Second, even when Retrieval-augmented generation (RAG)~\cite{lewis2020rag} injects retrieved evidence at query time to close this gap, the LLM's context window still imposes a hard limit on the number of candidates; injecting too many actually hurts because of attention decay over long contexts~\cite{liu2023lost}, so only the few candidates at the top truly influence the downstream answer.
Ranking quality, not merely recall, is therefore the key bottleneck of this retrieve-and-rerank pipeline; at the same time, general embedding models carry a systematic mean bias~\cite{ren2025r2} that misaligns distances on geoscience terminology, further amplifying the cost of mis-ranking.

To quantify how much \emph{agentic search} actually adds along this pipeline, the field needs both a trustable relevance signal and a verifiable evaluation infrastructure.
This paper operationalises these two terms into concrete engineering criteria.
\begin{itemize}
  \item \textbf{Trustable.} The relevance signal is taken from the datasets actually cited by peer-reviewed publications, which we call \emph{citation-grounded silver labels}. This signal supports comparative retrieval evaluation based on real scientific use, rather than the complete relevance ground truth of all datasets that could in principle apply to a research question. Because it comes from actual scientific use, the signal has high provenance.
  \item \textbf{Verifiable.} We release the benchmark and all citation-based metrics, so that any subsequent system can be quantitatively compared under the same protocol on the same evaluation set, rather than being asserted to work only through the authors' qualitative claims.
\end{itemize}
This paper studies the retrieve-and-rerank ranking layer between a research query and a dataset ranking, not end-to-end question answering. The term \emph{agentic} refers specifically to allowing the LLM to autonomously call external tools for evidence augmentation and terminology disambiguation at the rerank stage, not full-pipeline agentic operation.\footnote{We follow the \emph{agentic} definition of \citet{anthropic2024agents}, i.e.\ agent $=$ LLM $+$ autonomous tool use.}

Concretely, starting from the \texttt{USES\_DATASET} edges of the NASA Earth Observation Knowledge Graph (NASA EO-KG)~\cite{sun2025research}, we derive \sysname, a large-scale benchmark for geoscience dataset retrieval containing 47{,}654 query--dataset pairs.
On top of \sysname, we propose a retrieval correction suite (\nnname neural-score correction, a fine-tuned sentence transformer, and a convex combination with BM25). For the rerank stage, we contrast a single-shot LLM rerank prompt against an agentic harness that prepends a five-step web + arXiv research routine and grants autonomous tool access, holding the model, candidate set, and output contract fixed.
These two tools are used to supplement query-relevant paper and dataset evidence and to disambiguate ambiguous candidates, not to replace candidate recall.
The experiments give consistent evidence along two independent axes (Section~\ref{sec:experiments}). The retrieval suite lifts both R@10 and \MRR by more than $5\times$ over the unadapted cosine baseline; on the same LLM, enabling the agentic harness (a five-step web + arXiv research routine with autonomous tool calls) yields a directional \MAP/\MRR gain on the $N{=}200$ stratified test subset for the two LLMs we evaluate in this mode.
Compared with the closest prior frontier work, \citet{terrenzi2026agentic}, built on a fixed pre-built index, we instead use live web and arXiv tool calls, with the detailed distinction given in Section~\ref{sec:related}.

\noindent\textbf{Contributions.}
\begin{itemize}
  \item \textbf{\sysname benchmark.} 47{,}654 query--dataset positive pairs (21{,}272 citation-grounded queries, 38k/9k pairs for training/test) over a corpus of 8{,}058 NASA CMR datasets. To our knowledge in geoscience dataset retrieval, this is the \emph{first benchmark at this scale that supports supervised training}.
  \item \textbf{Retrieval suite.} \nnname neural-score correction, a fine-tuned sentence transformer, and BM25 fused through convex combination; the suite lifts both R@10 and \MRR by more than $5\times$ over the unadapted cosine baseline.
  \item \textbf{Controlled agentic-vs-LLM-rerank comparison.} On the same model, candidate set, and evaluation protocol, we contrast a single-shot LLM rerank prompt against an agentic harness that prepends a five-step web + arXiv research routine and grants autonomous tool access. The harness yields a directional \MAP/\MRR gain on the $N{=}200$ stratified subset for both vendors we evaluate (Opus 4.7 and DeepSeek v4 pro); the routine and tool access are coupled by design, with each routine step naming a specific tool, so single-shot LLM rerank is the natural control. This is explicitly distinguished from the pre-built-index approach of \citet{terrenzi2026agentic}.
\end{itemize}

The remainder of this paper is organised as follows. Section~\ref{sec:related} reviews related work; Section~\ref{sec:benchmark} presents the construction of \sysname; Section~\ref{sec:metrics} defines the evaluation metrics; Section~\ref{sec:methods} describes the method; Section~\ref{sec:experiments} reports the experiments; Section~\ref{sec:limitations} discusses limitations; and Section~\ref{sec:conclusion} concludes.


\section{Related Work}
\label{sec:related}

\paragraph{Geoscience LLM and multi-agent systems.}
LLMs have shown strong gains on individual downstream geoscience tasks~\cite{yi2025wdm, li2026hydroagent, yan2026aiagent, liu2025bams}, and autonomous multi-agent systems have demonstrated scalable scientific workflows across disciplines~\cite{ren2026autonomousresearch, sun2026agon, du2026autonumerics}.
Bringing the two together in geoscience and climate, AutoClimDS~\cite{jaber2025autoclimds} and PANGAEA-GPT~\cite{pantiukhin2026pangaea} build end-to-end agentic workflows spanning data discovery, analysis, and visualisation, positioning themselves as \emph{autonomous scientists} rather than retrieval systems; \citet{pantiukhin2025perspective} earlier sketched such uses of multi-agent systems on geoscience data in a Perspective.
Together with longer-running data-discovery research (geographic information retrieval, ontology-driven dataset discovery, and operational systems such as NASA Earthdata Search and CMR), this forms the domain context for our work.
It differs from ours in two respects.
First, these systems treat search as one component of a larger workflow or stay at metadata matching, whereas we study LLM-based search and rerank as an object in itself.
Second, they report no quantitative evaluation or only coarse scores from small LLM-as-a-judge runs, whereas we run a controlled comparison on 21k citation-grounded queries.

\paragraph{Knowledge graphs in geospatial data discovery.}
\citet{liu2026geo} propose a KG-driven multi-agent framework for data discovery across geospatial catalogs under heterogeneous metadata standards (the SpatioTemporal Asset Catalog, STAC; the Federal Geographic Data Committee Content Standard for Digital Geospatial Metadata, FGDC CSDGM; Data.gov; and others).
Their system uses a live Neo4j knowledge graph (264k datasets, 2.8M entities, 12M relations) as runtime infrastructure, where intent-parsing, graph-retrieval, and answer-synthesis agents collaborate at inference time. Evaluation uses 100 queries manually scored by two PhD students (0--3) on normalised discounted cumulative gain (NDCG@10), Recall@20, and a task-specific EIMR score.
Such work treats the KG as runtime infrastructure: every query's retrieval and ranking relies on its entities and relations.
We use the KG in a structurally different way, reading its \texttt{USES\_DATASET} edges once at benchmark construction to extract (publication, dataset) citation relations as silver labels. At runtime we query the graph only to fetch dataset records, never to traverse it: ranking comes from BM25 and dense scoring over text fields, and agentic rerank calls web and arXiv tools.
In effect we repurpose a KG built for publication--dataset linking as the source for an offline benchmark, which therefore inherits its coverage and curation bias---a boundary we state in Section~\ref{sec:limitations}.

\paragraph{Dataset retrieval benchmarks in information retrieval.}
The information retrieval (IR) community has built benchmarks and test collections for dataset retrieval: NTCIR-15 Data Search~\cite{kato2021ntcir}, ACORDAR~\cite{lin2022acordar}, bioCADDIE~\cite{cohen2017biocaddie}, a test collection~\citep{kolyada2025testcollection}, and DSEBench~\cite{shi2025dsebench}.
Most target general Web, Resource Description Framework (RDF), or biomedical data with hand-written or keyword queries, at scales from tens to thousands.
Industrial search engines and publication--dataset linking infrastructure (Google Dataset Search, DataCite, OpenAIRE, re3data, Scholix) target index scale and cross-platform citation graphs rather than fine-grained relevance evaluation, and work on dataset recommendation and scholarly data discovery likewise differs from the ad-hoc retrieval task studied here.
On the agentic and citation-driven side, FinAgentBench~\cite{choi2025finagentbench} builds the first agentic retrieval benchmark on financial-report passages, and \citet{tan2026citation} demonstrate a retrieval pattern based on citation context for multi-disciplinary computer science (CS) dataset discovery, close in spirit to our citation-grounded benchmark but differing in domain and ground-truth construction.
We differ from these in domain specificity, scale, and the source of the ground-truth (GT) signal: through the NASA EO-KG publication--dataset citation graph we scale to 21k geoscience queries, switching the GT signal from manual scoring or citation-context heuristics to the datasets a paper explicitly cites.
\sysname extends our own earlier pilot, ReSearch~\cite{sun2025research}, which introduced the GT-metrics protocol (Section~\ref{sec:metrics}).
Section~\ref{sec:limitations} discusses the recall-axis coverage limit of this use-based signal; we do not claim it equals every dataset applicable to the research question.

\paragraph{LLM-based and agentic reranking.}
Our agentic stage is a reranking stage, and LLM-based reranking has become an active direction in IR.
Representative works such as RankGPT~\cite{sun2023rankgpt}, listwise zero-shot reranking~\cite{ma2023listwise}, RankZephyr~\cite{pradeep2023rankzephyr}, and Rank-K~\cite{yang2025rankk} demonstrate zero-shot effectiveness on general corpora, in all cases restricting LLM reasoning to candidate text inlined in the prompt.
Supervised cross-encoder rerankers~\cite{nogueira2019passage} remain the standard non-LLM control for this stage, and we fine-tune one for comparison in Section~\ref{sec:ablation-model-sweep}.
Tool-augmented retrieval and QA agents such as ReAct~\cite{yao2022react}, Toolformer~\cite{schick2023toolformer}, and web-augmented QA introduce external knowledge access in open-domain multi-step reasoning.
Hypothetical-document and query-expansion methods such as HyDE~\cite{gao2022precise} evaluate primarily on open-domain QA and passage retrieval, which differ substantially from the fine-grained relevance ranking required in dataset retrieval.
On the dataset-retrieval side, \citet{terrenzi2026agentic} propose a reference architecture for agentic hybrid retrieval (BM25 + dense + Reciprocal Rank Fusion (RRF) + LLM agent plan/eval/rerank), but do not target geoscience or run a same-model controlled comparison of LLM against agentic rerank.
We differ in two ways.
First, we bring autonomous external tool use (web search and arXiv lookup) into the rerank stage of dataset retrieval, a combination not previously studied systematically.
Second, under the same LLM, same candidate set, and same output contract, we contrast a single-shot LLM rerank prompt against an agentic harness that prepends a five-step research routine with tool access, giving a same-model harness-vs-prompt comparison.

\section{Benchmark}
\label{sec:benchmark}

\subsection{Data Source and Construction}
\label{sec:bench-construction}

\sysname is a large-scale benchmark for geoscience dataset retrieval derived from
peer-reviewed NASA GES DISC publications.\footnote{%
  \sysname is publicly released at
  \url{https://huggingface.co/datasets/HamiltonMYu/NASA-EO-Bench}.}
Each query is grounded in an actual scientific paper.
From each paper's abstract, we prompt an LLM to generate a query phrased as a geoscience domain expert might naturally ask, simulating the retrieval intent of a future user of our agent, while the datasets explicitly cited by that paper serve as ground-truth answers.
This construction ensures that queries reflect genuine, diverse scientific information needs, grounded in real research practice rather than manually crafted prompts.

Concretely, we start from the
\emph{NASA Earth Observation Knowledge Graph}
(NASA EO-KG)~\cite{sun2025research},\footnote{%
  \url{https://huggingface.co/datasets/nasa-gesdisc/nasa-eo-knowledge-graph}}
which links publications to the NASA CMR datasets they cite.
We select the top 10{,}636 publications by citation count and generate two task-based queries per publication of the form \emph{``I want to \ldots''} using an LLM conditioned on the paper abstract (the full prompt is in the dataset card released with \sysname); the two queries are independent samples from the same prompt, intended to diversify the surface phrasing of the same underlying research intent. We retain only queries whose ground-truth answer set is non-empty after resolving against the CMR dataset corpus.
The resulting benchmark contains \textbf{47{,}654 query--dataset pairs} (21{,}272 task-based queries; mean 2.24 cited datasets per query, median 2) split into:

\begin{itemize}
  \item \textbf{Training set.} 38{,}078 pairs (17{,}038 queries).
  \item \textbf{Test set.} 9{,}576 pairs (4{,}234 queries).
\end{itemize}

\noindent
The 17{,}038 / 4{,}234 split is \emph{publication-level}: both queries from a given publication are kept in the same split, so no publication appears in both training and test, ruling out sibling-query leakage from the source-paper level. The split is stratified by \texttt{cited\_by\_count} so that the citation-count distribution is matched between training and test, avoiding domain shift across citation tiers.

\noindent
The dataset corpus consists of 8{,}058 NASA CMR datasets, each represented by its
\texttt{shortName}, \texttt{longName}, and free-text abstract.

\subsection{Comparison with Existing Benchmarks}
\label{sec:bench-comparison}

Table~\ref{tab:bench-compare} compares \sysname to prior geoscience retrieval benchmarks.
The ReSearch benchmark~\cite{sun2025research} contains 28 task-based queries from 14 publications, which is insufficient for stable metric estimation.
\sysname is roughly $\mathbf{760\times}$ larger in query count and provides a training
split that enables supervised model development.

\begin{table}[t]
  \centering
  \caption{Benchmark comparison.}
  \label{tab:bench-compare}
      \Description{Two-row table comparing benchmark sizes by query count. ReSearch has 28 queries, all in test, with no training split. NASA-EO-Bench has 21,272 queries, 17,038 for training and 4,234 for test.}
  \footnotesize
  \begin{tabular}{lrrr}
    \toprule
    \textbf{Benchmark} & \textbf{\#Queries} & \textbf{Train} & \textbf{Test} \\
    \midrule
    ReSearch~\cite{sun2025research} & 28       & --    & 28    \\
    \sysname (ours)                 & 21{,}272 & 17{,}038 & 4{,}234 \\
    \bottomrule
  \end{tabular}
\end{table}


\section{Evaluation Metrics}
\label{sec:metrics}

Following~\citet{sun2025research}, we evaluate retrieval quality with two complementary metric families, and we refer to them throughout the paper by the short names introduced here.
\emph{GT metrics} are the ground-truth (GT) citation-based measures $\RK{K}$, \MAP, and \MRR (Equations~\eqref{eq:rk}--\eqref{eq:ap}), whose positives are the datasets actually cited by the source publication; they quantify \emph{citation coverage}.
\emph{Judge metrics} are the LLM-as-a-Judge precision measures $\PK{K}$ (Equation~\eqref{eq:pk}), whose positives are decided independently by an LLM judge; they quantify \emph{semantic precision}.

\paragraph{Recall@$K$ (R@$K$)}
For a query $q$ with ground-truth dataset set $\mathcal{G}_q$, and a ranked list $\mathcal{R}_q$ of top-$K$ retrieved datasets, Recall@$K$ is:
\begin{equation}
  \RK{K}(q) \;=\; \frac{|\mathcal{R}_q^{(K)} \cap \mathcal{G}_q|}{|\mathcal{G}_q|}
  \label{eq:rk}
\end{equation}
where $\mathcal{R}_q^{(K)}$ denotes the top-$K$ results.
The macro-averaged score $\RK{K} = \frac{1}{|\mathcal{Q}|}\sum_{q}\RK{K}(q)$ is reported across query set $\mathcal{Q}$.
We report $K \in \{10, 20, 100\}$; R@100 measures the upper bound available to precision-oriented re-ranking pipelines.

\paragraph{Mean Reciprocal Rank (\MRR)}
\MRR rewards retrieving \emph{any} relevant dataset early in the ranking:
\begin{equation}
  \MRR \;=\; \frac{1}{|\mathcal{Q}|} \sum_{q \in \mathcal{Q}} \frac{1}{\mathrm{rank}_q}
  \label{eq:mrr}
\end{equation}
where $\mathrm{rank}_q$ is the position of the first relevant dataset in the ranked list for query $q$.
In our agentic pipeline, \MRR is the most operationally critical metric: the LLM reranker operates on a fixed top-$K$ window, so a relevant dataset that ranks late may never be seen or acted upon.

\paragraph{Mean Average Precision (\MAP)}
\MAP captures the quality of the entire ranked list:
\begin{equation}
  \mathrm{AP}(q) \;=\; \frac{1}{|\mathcal{G}_q|}
    \sum_{k=1}^{|\mathcal{R}_q|} P_q(k) \cdot \mathrm{rel}_q(k)
  \label{eq:ap}
\end{equation}
where $P_q(k)$ is precision at cut-off $k$, and $\mathrm{rel}_q(k) \in \{0,1\}$ indicates whether the $k$-th retrieved item is in $\mathcal{G}_q$. $\MAP = \frac{1}{|\mathcal{Q}|}\sum_q \mathrm{AP}(q)$.

\paragraph{LLM-as-a-Judge Precision@$K$ ($\PK{K}$).}
Citation-based metrics are inherently \emph{recall-biased}, in that a dataset that is scientifically relevant to a query but not cited by the source paper is penalised.
To complement GT metrics, we employ an LLM judge that independently decides, for each (query, dataset) pair, whether the dataset ``\emph{measures the same physical variable or phenomenon as the query requires}.''
Concretely, we serve \texttt{Qwen3.6-35B-A3B} (a mixture-of-experts, MoE, model with 3B active parameters) locally via vLLM and prompt it with the query text and the dataset description (\texttt{shortName} + abstract).
The judge returns a binary YES/NO decision.
Precision@$K$ is then:
\begin{equation}
  \PK{K}(q) \;=\; \frac{1}{K} \sum_{k=1}^{K} \mathbf{1}[\text{judge}(q, d_k) = \text{YES}]
  \label{eq:pk}
\end{equation}
macro-averaged across queries.
We report $K \in \{5, 10, 20\}$.


\section{Method}
\label{sec:methods}

\subsection{Problem Formulation and Design Overview}
\label{sec:agent-pipeline}

Figure~\ref{fig:pipeline} shows the full pipeline as a three-stage architecture.

\paragraph{Stage 1: NASA official tools.}
Many queries can be directly answered by NASA's existing ecosystem---\emph{Harmony} for data access and subsetting, \emph{SDE} for domain-knowledge and event-background lookup, and \emph{WorldView}/\emph{Giovanni} for visualisation.
The Router dispatches to these tools first; where any tool fully satisfies the request, the pipeline terminates here with provenance-annotated output, avoiding unnecessary retrieval.

\paragraph{Stage 2: Hybrid retrieval.}
When no official tool fully resolves the query, the pipeline falls back to dataset retrieval over the NASA CMR corpus.
Because dataset relevance may be expressed through exact scientific identifiers or through paraphrased research intent, we combine lexical and semantic evidence to cover both signals.

\paragraph{Stage 3: Agentic reranking.}
Because the highest-ranked candidates from Stage~2 may still contain ambiguities that corpus metadata alone cannot resolve---such as a dataset's actual coverage for a named event, or its common usage in the relevant literature---we apply bounded tool-grounded reranking only within the top-$K$ candidate window.
Confining reasoning to this window keeps cost proportional to candidate count while preserving the retriever's broader ranking.

The remainder of this section describes each component in turn: lexical anchoring (Section~\ref{sec:bm25}), task-adaptive semantic scoring (Section~\ref{sec:nn-ssc}), hybrid fusion (Section~\ref{sec:hybrid}), and the two reranking stages (Sections~\ref{sec:llm-rerank}--\ref{sec:agentic-rerank}).

\subsection{Lexical Anchoring}
\label{sec:bm25}

NASA dataset records are identified by mission names, instrument codes, product short names, and physical variable labels---exact tokens that carry high-precision relevance signal when they appear in a query.
We retain BM25~\cite{robertson1995okapi} as a lexical anchor: operating on term statistics, it scores these domain-specific tokens directly from corpus vocabulary without relying on any learned representation.
Its role is not to model semantic intent but to preserve discriminative surface-form evidence.
BM25 cannot recover datasets whose metadata paraphrases the same scientific need in different terminology, motivating the semantic component below.

\subsection{Task-Adaptive Semantic Scoring}
\label{sec:nn-ssc}

Textual similarity is not the same as dataset suitability.
A pretrained encoder trained on general text may score a dataset as similar to a query because it \emph{mentions} related concepts, even if it lacks the required variables or measurement type.
Domain shift further de-calibrates distances in a pair-specific manner: geoscience terminology is unevenly represented in general pretraining corpora, so the mismatch between query and dataset embeddings varies across pairs rather than shifting uniformly. A global linear correction such as mean-bias removal~\cite{ren2025r2} successfully removes the shared mean component of this bias; the residual, however, is pair-specific and non-uniform, motivating a non-linear per-pair adaptation.

Both failure modes call for supervised adaptation.
We use citation-grounded (query, dataset) pairs from the training split and mine hard negatives from the nearest neighbours of each query in the base embedding space, the candidates most likely to be falsely retrieved.
The training objective, for each positive $d^+ \in \mathcal{P}$ and hard-negative set $\mathcal{N}$, is to assign $d^+$ a higher relevance score than all $n \in \mathcal{N}$,
\begin{equation}
  \mathcal{L} \;=\; \frac{1}{|\mathcal{P}|}\sum_{p \in \mathcal{P}}
    -\log \frac{\exp(s_p)}
               {\exp(s_p) + \sum_{n \in \mathcal{N}} \exp(s_n)} ,
  \label{eq:supcon}
\end{equation}
where $s_p, s_n \in (0,1)$ are the scorer's sigmoid outputs for the positive and each negative.
Equation~\eqref{eq:supcon} is the standard InfoNCE (softmax cross-entropy) ranking loss, applied once per positive: each $d^+$ forms its own softmax against the whole negative set $\mathcal{N}$, and the per-positive terms are then averaged.
Positives therefore never appear in each other's denominator.
This is the appropriate form for our setting, where a query carries $2.24$ ground-truth datasets on average: a single list-wise softmax over all candidates would place the co-cited datasets of one publication in competition, penalising each for the others' scores.
We consider two implementation variants.

\paragraph{Pairwise score correction (NN-SSC, \nnname).}
The encoder is kept frozen; a lightweight multi-layer perceptron (MLP) $f_\theta: \mathbb{R}^{2D}\!\to\!\mathbb{R}$ takes the concatenated query--dataset embedding pair and outputs a scalar relevance score.
Because relevance correction is treated as a pair-specific function, different query--dataset pairs can receive different corrections rather than a single global transformation.

\paragraph{Encoder fine-tuning.}\label{sec:ftst}
Alternatively, we fine-tune the encoder on (query, positive, hard negative) triples under the same ranking objective, directly adjusting the embedding geometry so that relevant datasets move closer to the query representation.
This improves all cosine-based downstream components without changing their architectures.

Both variants produce a semantic score $\hat{s}_\text{sem}$ that feeds into the fusion step.

\subsection{Hybrid Lexical--Semantic Fusion}
\label{sec:hybrid}

BM25 and neural semantic scores operate in incompatible score spaces: BM25 outputs unbounded term-frequency statistics while \nnname produces values in $(0,1)$ via a sigmoid.
We apply min-max normalisation to map each score sequence to the unit interval before mixing, yielding a convex combination:
\begin{equation}
  s_{\text{hybrid}}(q, d) \;=\; \alpha \cdot \hat{s}_{\text{NN}}(q,d)
                              \;+\; (1-\alpha)\cdot \hat{s}_{\text{BM25}}(q,d)
  \label{eq:hybrid}
\end{equation}
where $\hat{\cdot}$ denotes min-max normalisation over all datasets for query $q$, and $\alpha \in [0,1]$ trades off lexical precision against semantic recall.
\citet{bruch2024analysis} ground this geometry: in the 2D lexical--semantic score space, relevant documents cluster such that a linear boundary separates them from irrelevant ones, making the convex combination the natural scoring function that outperforms Reciprocal Rank Fusion in both in-domain and out-of-domain settings.

\paragraph{Modularity.}
The fusion layer is agnostic to the choice of semantic scorer: $\hat{s}_{\text{NN}}$ can be replaced by cosine similarity, a fine-tuned encoder score, or any future \nnname variant without altering the mixing logic, allowing independent improvement of each component.

\paragraph{Setting $\alpha$ analytically.}
We compute $\alpha$ from the standalone retrieval ability of the two scorers.
Let $\pi_{\ell}$ and $\pi_n$ denote the standalone GT-metric performance (averaged over R@10, R@20, R@100, \MAP, and \MRR) of BM25 and the neural scorer, respectively, each measured on the \emph{training split}; the coefficient is
\begin{equation}
  \alpha \;=\; \frac{\pi_n}{\pi_{\ell} + \pi_n},
  \label{eq:alpha}
\end{equation}
so that the fusion weight equals the neural scorer's share of the two standalone performances.
Appendix~\ref{app:alpha-derivation} motivates this relative-performance form under a squared-loss surrogate: with fixed component scores, the loss is a convex quadratic in $\alpha$, and weakly correlated residual errors lead to the same inverse-error weighting rule.
With this rule, $\alpha = 0.5$ when the two scorers are equally accurate, $\alpha \!\to\! 1$ when the lexical signal is uninformative, and $\alpha \!\to\! 0$ when the neural signal is.
The test split is reserved for final reporting; concrete values of $\alpha$ for each hybrid configuration are reported in Section~\ref{sec:experiments}.

\subsection{LLM Reranking}
\label{sec:llm-rerank}

The hybrid retrieval stage produces a ranking based on combined lexical and semantic scores.
However, a retriever trained on publication citations carries two structural risks: popular datasets tend to be over-represented in citation data (Matthew effect), and cited datasets are not the only relevant ones, as many suitable alternatives are simply never cited by any training paper (Section~\ref{sec:limitations}).
Because an LLM judges dataset suitability from general knowledge rather than citation co-occurrence patterns, it can partially compensate for both biases: given a query and a small set of top candidates, it reasons about research intent, weighs candidates against each other, and surfaces mismatches that lexical overlap and embedding distance alone cannot capture.

Concretely, we take the top candidates from the retrieval output, fill them into a structured prompt together with the query, and submit the prompt to the LLM for reranking.
The LLM returns an ordered list of candidate ids, which is then composed in three segments by Algorithm~\ref{alg:agentic}: LLM-ranked candidates first, unmentioned top-$K$ candidates in their original order, and items outside the window unchanged.
The full prompt is given in Appendix~\ref{app:rerank-prompt}.

\paragraph{Flexibility.}
LLM reranking requires no task-specific training; adapting to a new task requires only editing the prompt's sorting criteria.
The same prompt format and output contract work with any instruction-following LLM, enabling fair cross-model comparison under a fixed protocol (Section~\ref{sec:ablation-rerank}).

\subsection{Agentic Reranking with Autonomous Tool Use}
\label{sec:agentic-rerank}

LLM reranking has an inherent limitation: the model can only access the query and candidate text inlined in the prompt, leaving information such as ambiguities in candidate descriptions, detailed usage information absent from the dataset abstract, or the context of event-specific queries (e.g., a named flood or wildfire that requires external knowledge) simply unavailable.
Without access to such external context, the gains from reranking are fundamentally bounded.

Agentic reranking addresses this by equipping the LLM with autonomous tool access during the reranking step.
Before producing the final ranking, the model may invoke web search for up-to-date context and arXiv paper lookup to identify datasets commonly used for a given research task.
The broader system design also includes SDE search as an optional NASA-tool integration, although it is not exercised in the controlled experiments reported here.
All tools are called at the model's own discretion; Algorithm~\ref{alg:agentic} summarises the full procedure.

\begin{algorithm}[t]
  \small
  \begin{algorithmic}[1]
    \REQUIRE query $q$, retrieval-ranked list $R = [r_1, \ldots, r_N]$, window size $K$
    \ENSURE reranked list $R'$
    \STATE $C \gets [r_1, \ldots, r_K]$
      \COMMENT{top-$K$ candidates}
    \STATE LLM autonomously invokes tools (web search and arXiv paper lookup in the experiments here) and reasons over $q$ and $C$;
            outputs an ordered list $L = [a_1, \ldots, a_m]$, where each $a_j \in \{1,\ldots,K\}$ and all distinct
    \STATE $R' \gets [r_{a_1}, \ldots, r_{a_m}]$
            \COMMENT{LLM-ranked order}
            \newline\hspace*{\algorithmicindent} ${}+ [\,r_i \;\text{for}\ i = 1..K,\ i \notin L\,]$
            \COMMENT{unmentioned, original order}
            \newline\hspace*{\algorithmicindent} ${}+ [r_{K+1}, \ldots, r_N]$
            \COMMENT{tail unchanged}
    \RETURN $R'$
  \end{algorithmic}
  \caption{Agentic Reranking}
  \label{alg:agentic}
\end{algorithm}

LLM reranking (Section~\ref{sec:llm-rerank}) is a degenerate version of this algorithm in which Step~2 is replaced by ranking directly over the inlined candidate text with no tool calls; the candidate XML format, sorting instructions, and JSON output contract are identical in both settings.
This yields a same-model comparison (same model, same candidates, same evaluation protocol) in which the agentic side adds both a prepended five-step research routine and tool access. The routine is the natural-language protocol that drives the tool calls, so the two are inseparable by design and single-shot LLM rerank is the natural control (Section~\ref{sec:ablation-agentic}).


\section{Experiments}
\label{sec:experiments}

\subsection{Experimental Setup}

\paragraph{Datasets and evaluation protocol.}
Benchmark construction is detailed in Section~\ref{sec:benchmark}, and evaluation metrics are defined in Section~\ref{sec:metrics}.
This section supplies implementation details that are orthogonal to the benchmark design but necessary for reproducing the results.

All embedding-based methods share \path|nasa-smd-ibm-st-v2|~\cite{nasa-smd-ibm-st-v2},
a 768-dimensional sentence transformer adapted on a NASA scientific question-answering corpus.
Its prior exposure to NASA terminology, instrument names, and geoscience concepts mitigates domain shift relative to general-purpose encoders before task-specific correction.
Dataset embeddings are pre-computed and cached.

For all embedding-based methods, the textual representation of each dataset is formed by concatenating its
\texttt{shortName}, \texttt{longName}, and \texttt{abstract} (truncated to 512 characters).
The concatenation is passed through the shared backbone encoder described above,
L2-normalised, and then used either for cosine similarity or as input to NN-SSC.
For BM25, in addition to the three fields above, we also index the DOI, the DAAC source identifier, and the temporal coverage,
to cover queries that mention specific satellite missions or data centers.

The fusion coefficient $\alpha$ is set analytically from scorer performances via Equation~\eqref{eq:alpha}, avoiding both test-set and validation-set tuning; the test split is strictly used for reporting results.

\paragraph{Compared methods.}
We evaluate the following seven methods on the same test set, which are also all rows of the main table (Table~\ref{tab:main}).

\begin{itemize}

  \item \textbf{BM25}~\cite{robertson1995okapi}.
    An inverted index of dataset records is built with Okapi BM25 ($k_1{=}1.5$, $b{=}0.75$).
    Indexed fields include shortName, longName, abstract, and the remaining metadata fields.
    No training is needed; lexical retrieval runs directly on the test queries.

  \item \textbf{Cosine (base)}.
      The pre-trained NASA-SMD-IBM-ST-v2 encodes both queries and dataset records~\cite{nasa-smd-ibm-st-v2},
    and the cosine similarity after L2 normalisation is used as the ranking score.
    No learnable parameters are introduced.

  \item \textbf{Cosine (fine-tuned)}.
    Following Section~\ref{sec:ftst}, we fine-tune the bi-encoder for 10 epochs with
    MultipleNegativesRankingLoss ($\tau{=}0.07$), five cosine-KNN hard negatives per positive,
    and in-batch negatives (batch size 256, AdamW, learning rate $2{\times}10^{-5}$, warmup 0.1, BF16).

  \item \textbf{NN-SSC}.
    The neural semantic score-correction method proposed in this paper (§\ref{sec:nn-ssc}).
    The input is the concatenation of the L2-normalised query embedding and dataset embedding (1536 dimensions),
    passed through a three-layer fully connected network ($1536{\to}256{\to}256{\to}1$, ReLU activations in the hidden layers,
    Sigmoid at the output) that produces a relevance score.
    \nnname is trained with the InfoNCE objective in Equation~\eqref{eq:supcon}, using 10 KNN hard and 10 random negatives per positive for 1000 epochs
    (batch size 64, Adam, learning rate $10^{-4}$, seed 42).

  \item \textbf{Cosine (fine-tuned) + BM25 (Hybrid)}.
    The fine-tuned cosine score and the BM25 score are each min-max normalised to $[0,1]$
    and linearly fused as $\hat{s}_{\text{hybrid}} = \alpha \cdot \hat{s}_{\text{cosine}} + (1-\alpha) \cdot \hat{s}_{\text{BM25}}$,
    with $\alpha$ set analytically by Equation~\eqref{eq:alpha}, giving $\pi_n/(\pi_{\ell}+\pi_n) \approx 0.75$.

  \item \textbf{NN-SSC + BM25 (Hybrid)}.
    The NN-SSC score and the BM25 score are each min-max normalised and then linearly fused,
    with $\alpha$ set analytically by Equation~\eqref{eq:alpha}, giving $\pi_n/(\pi_{\ell}+\pi_n) \approx 0.76$.

  \item \textbf{Cosine (fine-tuned) + BM25 (Hybrid) + R2}.
    A training-free post-processing variant of Cosine (fine-tuned) + BM25 (Hybrid).
    After encoding, R2~\cite{ren2025r2} projects both document and query embeddings away from the unit mean direction $\hat{\mu}$ of the combined document and training-query corpus, followed by L2 normalisation.
    These embeddings replace the standard cosine scores; BM25 fusion uses the same $\alpha$ as the fine-tuned cosine hybrid.

\end{itemize}

\paragraph{Evaluation metrics.}
We use the GT metrics and LLM Judge P@$K$ defined in Section~\ref{sec:metrics}, with \MAP and \MRR computed over the top-100 candidates.
While GT metrics reflect citation coverage and Judge metrics semantic precision, we use both for a
comprehensive evaluation.
Rerank gains are tested with a paired bootstrap over per-query metric differences ($10{,}000$ resamples):
the 95\% CI is the $2.5$/$97.5$ percentile of the resampled mean difference, and the two-sided $p$ is
twice the smaller tail mass beyond zero.

\paragraph{Reproducibility.}
NN-SSC and the fine-tuning experiments run on an NVIDIA B200 GPU with random seed 42.
LLM API requests (Sections~\ref{sec:ablation-rerank} and~\ref{sec:ablation-agentic}) target pinned snapshots: \path|gpt-5.5-2026-04-23|, \path|gpt-5.4-2026-03-05|, \path|o4-mini-2025-04-16|, \path|claude-opus-4-7|, and \path|deepseek-v4-pro|. Only \path|gpt-5.4| accepts a temperature override and is run at temperature $0$; the reasoning and newest-flagship models reject it and run at their provider default.
The \sysname dataset card documents the benchmark-construction procedure and the query-generation prompt; the fixed LLM and agentic rerank prompt templates are in Appendix~\ref{app:rerank-prompt}.

\subsection{Main Results}
\label{sec:results}

\begin{table*}[t]
  \centering
  \caption{Retrieval performance on \sysname test set.}
  \label{tab:main}
    \Description{Retrieval results for seven methods on the NASA-EO-Bench test set. Columns give ground-truth Recall at 10, 20 and 100, MAP, MRR, and LLM Judge Precision at 5, 10 and 20. Rows run from the unadapted cosine baseline and BM25 through fine-tuned cosine, the two hybrids, NN-SSC alone, and NN-SSC fused with BM25, which leads on the ground-truth metrics.}
  \footnotesize
  \setlength{\tabcolsep}{2pt}
    \begin{tabular}{lcccccccc}
      \toprule
      \multirow{2}{*}{\textbf{Method}}
        & \multicolumn{3}{c}{\textbf{GT Recall@K}}
        & \multirow{2}{*}{\textbf{\MAP}}
        & \multirow{2}{*}{\textbf{\MRR}}
        & \multicolumn{3}{c}{\textbf{LLM Judge P@K}} \\
      \cmidrule(lr){2-4} \cmidrule(lr){7-9}
        & R@10 & R@20 & R@100
        & & & P@5 & P@10 & P@20 \\
      \midrule
      Cosine (base)
        & 0.0755 & 0.1147 & 0.2629
        & 0.0402 & 0.0538
        & 0.1406 & 0.1285 & 0.1140 \\
      Cosine (fine-tuned)
        & 0.3632 & 0.4591 & 0.6962
        & 0.2117 & 0.2423
        & 0.3033 & 0.2678 & 0.2279 \\
      BM25
        & 0.1083 & 0.1529 & 0.3026
        & 0.0643 & 0.0805
        & 0.1848 & 0.1670 & 0.1431 \\
      Cosine (fine-tuned) + BM25 (Hybrid)
        & 0.3692 & 0.4725 & 0.7091
        & 0.2170 & 0.2522
        & 0.3362 & 0.2962 & 0.2492 \\
      Cosine (fine-tuned) + BM25 (Hybrid) + R2
        & 0.3710 & 0.4735 & 0.7125
        & 0.2183 & 0.2540
        & \textbf{0.3367} & \textbf{0.2966} & \textbf{0.2498} \\
      \nnname
        & 0.3614 & 0.4912 & 0.7700
        & 0.2182 & 0.2480
        & 0.1800 & 0.1588 & 0.1359 \\
      \nnname + BM25 (Hybrid)
        & \textbf{0.4275} & \textbf{0.5530} & \textbf{0.8023}
        & \textbf{0.2495} & \textbf{0.2918}
        & 0.2915 & 0.2517 & 0.2040 \\
      \bottomrule
    \end{tabular}
  \par
  \parbox{\textwidth}{\footnotesize\raggedright
    R@$K$, \MAP, and \MRR are ground-truth citation-based; P@$K$ is
    LLM-as-a-Judge (\texttt{Qwen3.6-35B-A3B}, physical-variable matching
    prompt). Best in each column in \textbf{bold}.}
\end{table*}

\begin{table}[t]
  \centering
  \caption{Complexity comparison of \nnname vs.\ the fine-tuned encoder.}
  \label{tab:complexity}
    \Description{Five-row comparison of NN-SSC against the fine-tuned encoder: 460K trainable parameters versus about 125M, a shared frozen 125M encoder versus none, about 125M total parameters on both sides, backpropagation through the MLP only versus the full encoder, and inference over cached embeddings plus an MLP versus base cosine.}
  \footnotesize
  \setlength{\tabcolsep}{4pt}
    \begin{tabular}{@{}lcc@{}}
      \toprule
      & \textbf{\nnname} & \textbf{Cosine (fine-tuned)} \\
      \midrule
      Trainable params         & 460K               & $\sim$125M \\
      Frozen shared encoder    & 125M               & ---  \\
      Total params             & $\sim$125M         & $\sim$125M \\
      Backprop through         & MLP only           & Full encoder \\
      Inference cost           & Cached emb.\ + MLP & Base cosine \\
      \bottomrule
    \end{tabular}
\end{table}

Table~\ref{tab:main} reports the results of all methods on the \sysname test set.

\paragraph{Domain shift is the main performance bottleneck.}
Cosine (base) sits below BM25 on R@10, showing that an unadapted pre-trained embedding gives an unreliable relevance signal in the geoscience domain and that lexical matching already encodes an independent, effective signal.
Every hybrid row in the table beats its standalone neural component, so BM25's complementary contribution is stable across neural scorers.
The Cosine (fine-tuned) + BM25 + R2 row further corroborates this: mean-bias removal~\cite{ren2025r2}, a training-free global linear correction, still yields a small but consistent gain even on top of a fine-tuned encoder, confirming that residual domain bias is not fully resolved by encoder adaptation alone.
\nnname's substantially larger gains through non-linear per-pair supervision show that the domain mismatch is pair-specific and better addressed by learned adaptation than by a uniform linear shift.

\paragraph{Both metric families confirm real accuracy gains.}
Each shows a clear lift over the base cosine baseline, and the GT-best method \nnname\,+\,BM25 more than doubles Judge P@5 over the unadapted cosine baseline.
Since our scorers are citation-supervised, a GT gain alone could mean they merely surface cited-but-irrelevant datasets, but the parallel gain in Judge P@5 reduces this concern.
Directional agreement does not, however, mean the two families pick the same best method: the fine-tuned cosine hybrids attain the highest Judge P@5 ($0.3367$, with R2), while \nnname\,+\,BM25 leads on all GT metrics.
This partly reflects the Qwen judge favouring results aligned with general semantic space, a direction different from \nnname's citation-cooccurrence supervision signal.
Either method can fill the semantic-correction slot of our pipeline, so what distinguishes them is parameter cost rather than performance ceiling (Table~\ref{tab:complexity}).

\paragraph{Complexity comparison.}%
\label{sec:results-complexity}
\nnname adds only $460$K trainable parameters on top of the same frozen $125$M backbone that all cosine baselines already use, roughly $1/270$ of the fine-tuned encoder's trainable count, yet matches or exceeds the fine-tuned encoder on every GT metric except R@10, where the two are within $0.002$; the gap widens after fusion with BM25.
Because the backbone stays frozen and the MLP runs over cached embeddings, \nnname can be plugged on top of any pre-trained encoder without paying the cost of full encoder fine-tuning, making it especially valuable when full fine-tuning is infeasible and a general-purpose contribution beyond geoscience dataset retrieval.

\paragraph{Hybrid coefficient.}
\label{sec:ablation-alpha}
The mixing coefficient $\alpha$ is set analytically per Equation~\eqref{eq:alpha} of Section~\ref{sec:hybrid}, so no $\alpha$-sweep on the test or validation set is needed.
Dropping BM25 entirely (i.e.\ ranking by \nnname\ alone) loses both \MAP\ and \MRR, validating that BM25's exact lexical signal is indispensable for placing the most relevant results near the top.
\nnname\,+\,BM25 is the retrieval backbone for all subsequent experiments except the backbone ablation below.

\subsection{LLM Reranking}
\label{sec:ablation-rerank}

The retrieval-stage ordering is decided only by BM25's lexical signal and NN-SSC's embedding signal, leaving systematic relative-order errors among the top candidates.
We ask whether zero-shot LLM reranking can deliver stable gains on \sysname and how that gain relates to upstream retrieval strength.

We pick two retrieval backbones, Cosine (base)~+~BM25 and \nnname~+~BM25 (weak and strong retrieval ceilings), and apply one round of GPT-5.5 reranking on each.
Rerank experiments use subsets of the test set rather than all $4{,}234$ queries to keep LLM API cost tractable, stratified by source-publication citation count to preserve the long-tail distribution. All three rerank tables were first run on a common $N{=}200$ subset and agreed there; we then extended the model sweep (Table~\ref{tab:model-sweep}) to $N{=}600$ with paired-bootstrap CIs and $p$-values, which sharpen those conclusions rather than revise them. The backbone ablation (Table~\ref{tab:rerank-ablation}) and the agentic comparison (Table~\ref{tab:agentic-vs-api}) retain the original $N{=}200$ runs, the latter because tool calls raise per-query cost by an order of magnitude (Section~\ref{sec:limitations}); absolute scores are comparable only within a subset size.
For each query the LLM receives the top-10 retrieval candidates and returns a reordered top-10.

\begin{table}[t]
  \centering
  \caption{GPT-5.5 reranking on two retrieval backbones.}
  \label{tab:rerank-ablation}
    \Description{GPT-5.5 reranking applied to two retrieval backbones, weak (base cosine plus BM25) and strong (NN-SSC plus BM25). Columns are Recall at 10, 20 and 100, MAP and MRR. Recall is unchanged by reranking; both backbones gain in MAP and MRR.}
  \footnotesize
    \begin{tabular*}{\columnwidth}{@{\extracolsep{\fill}}lccccc@{}}
      \toprule
      \textbf{Setup} & \RK{10} & \RK{20} & \RK{100} & \MAP & \MRR \\
      \midrule
      Cosine (base) + BM25                    & 0.132 & 0.172 & 0.364 & 0.067 & 0.082 \\
      \quad+~GPT-5.5                          & 0.132 & 0.172 & 0.364 & \textbf{0.109} & \textbf{0.143} \\
      \midrule
      \nnname + BM25                          & 0.483 & 0.609 & 0.829 & 0.260 & 0.302 \\
      \quad+~GPT-5.5                          & 0.483 & 0.609 & 0.829 & \textbf{0.322} & \textbf{0.383} \\
      \bottomrule
    \end{tabular*}
  \par
  \parbox{\columnwidth}{\footnotesize\raggedright
    Rerank rows with higher \MAP/\MRR than their backbone baseline are set in
    \textbf{bold}. The no-rerank baseline is re-stated in
    Table~\ref{tab:agentic-vs-api}, which uses the same subset.}
\end{table}

Table~\ref{tab:rerank-ablation} shows that GPT-5.5 delivers a positive \MAP/\MRR lift on both backbones, with the shape of the gain varying inversely with upstream retrieval strength.
The weak backbone yields the larger relative gain (the correct answers are scattered across the top-10, giving the LLM more room to rerank); the strong backbone yields the larger absolute gain.
In both cases the LLM uses no retrieval-specific supervision and reranks purely from the candidate text inlined in the prompt.

The R@K columns are identical before and after reranking by construction.
The LLM only permutes candidates within the top-10 window and does not introduce new candidates, so $K{\ge}10$ recall is fully determined by the retrieval stage and the observed \MAP/\MRR gains attach to the LLM's intra-window reordering rather than to changes in the recall set.

\subsection{Reranker Model Comparison}
\label{sec:ablation-model-sweep}

Section~\ref{sec:ablation-rerank} established that LLM reranking is effective on \sysname.
We next check whether this gain is robust across LLMs and identify the strongest single-shot baseline for the agentic comparison in Section~\ref{sec:ablation-agentic}.

We fix the retrieval backbone to \nnname~+~BM25 and evaluate five LLMs on the $N{=}600$ subset and rerank protocol, spanning three vendors (OpenAI o4-mini / GPT-5.4 / GPT-5.5, Anthropic Claude Opus 4.7, DeepSeek v4 pro) and both reasoning and non-reasoning models.

\begin{table*}[t]
  \centering
  \caption{Reranker comparison.}
  \label{tab:model-sweep}
    \Description{Six rerankers on the NN-SSC plus BM25 backbone, each reported as MAP and MRR with the paired difference against no reranking, its 95 percent confidence interval, and a p-value. Rows are a fine-tuned cross-encoder followed by five zero-shot LLMs: o4-mini, GPT-5.4, DeepSeek v4 pro, GPT-5.5 and Claude Opus 4.7, which leads on both metrics.}
  \footnotesize
  \setlength{\tabcolsep}{4pt}
    \begin{tabular*}{\textwidth}{@{\extracolsep{\fill}}lcrcrcr@{}}
      \toprule
      & \multicolumn{3}{c}{\MAP} & \multicolumn{3}{c}{\MRR} \\
      \cmidrule(lr){2-4}\cmidrule(lr){5-7}
      \textbf{Setup} & score & $\Delta$ (95\% CI) & $p$ & score & $\Delta$ (95\% CI) & $p$ \\
      \midrule
      no rerank                   & 0.251 & --- & & 0.301 & --- & \\
      cross-encoder (ft)\textsuperscript{a} & 0.276 & $+$0.026 $[+0.004,+0.047]$ & 0.018 & 0.321 & $+$0.020 $[-0.008,+0.047]$ & 0.155 \\
      \multicolumn{7}{@{}l}{\emph{Zero-shot LLM rerankers}} \\
      \quad o4-mini               & 0.260 & $+$0.009 $[-0.010,+0.028]$ & 0.332 & 0.322 & $+$0.021 $[-0.003,+0.045]$ & 0.082 \\
      \quad GPT-5.4               & 0.274 & $+$0.024 $[+0.004,+0.044]$ & 0.021 & 0.332 & $+$0.031 $[+0.005,+0.057]$ & 0.022 \\
      \quad DeepSeek v4 pro       & 0.276 & $+$0.025 $[+0.005,+0.046]$ & 0.014 & 0.337 & $+$0.036 $[+0.010,+0.062]$ & 0.007 \\
      \quad GPT-5.5               & 0.276 & $+$0.025 $[+0.005,+0.046]$ & 0.014 & 0.339 & $+$0.038 $[+0.013,+0.063]$ & 0.003 \\
      \quad Claude Opus 4.7       & \textbf{0.294} & $+$\textbf{0.043} $[+0.023,+0.064]$ & ${<}0.001$ & \textbf{0.350} & $+$\textbf{0.048} $[+0.023,+0.074]$ & ${<}0.001$ \\
      \bottomrule
    \end{tabular*}
  \par
  \parbox{\textwidth}{\footnotesize\raggedright
    All rerankers share the same \nnname~+~BM25 stage-1 candidates and reorder the
    top-10, so R@$K$ is unchanged by construction. $\Delta$ is the paired difference
    against no rerank over the same 600 queries (paired bootstrap, 10{,}000 resamples); a gain
    counts as significant when $p < 0.05$.\par
    \textsuperscript{a}\,\texttt{bge-reranker-v2-m3} fine-tuned on \sysname\ with negatives drawn from
    the stage-1 top-100.}
\end{table*}

Four of the five LLMs improve significantly over the no-rerank baseline (Table~\ref{tab:model-sweep}), so the zero-shot rerank gain is not tied to a single model; only o4-mini fails to reach significance on either metric.
Claude Opus 4.7 takes both top \MAP and top \MRR and is therefore the strongest single-shot baseline on this benchmark, significantly ahead of GPT-5.5 on \MAP ($\Delta{=}{+}0.018$, $p{=}0.001$; on the smaller $N{=}200$ subset the two are within noise).
The one small reasoning model in the sweep, o4-mini, is the weakest, potentially indicating that capacity rather than reasoning budget sets the ceiling on this short-context fixed-format ranking task.
As a supervised reranking control we also fine-tune \texttt{bge-reranker-v2-m3}~\cite{chen2024bge,li2023making} on \sysname,
drawing negatives from the stage-1 top-100. This tuned cross-encoder matches the LLM rerankers
on \MAP\ but not on \MRR, where its gain is not significant --- the LLM advantage lies in
first-hit placement rather than overall ordering.
It is a competitive baseline, but a fragile one: under a poor fine-tuning recipe it ranks worse
than not reranking at all. The LLM rerankers carry no such risk --- they improve the ranking
zero-shot, and stand to keep improving as the models themselves do.
For the agentic experiments in Section~\ref{sec:ablation-agentic} we use Opus 4.7 and DeepSeek v4 pro instead; our in-house agent harness wraps the Claude and DeepSeek APIs natively at the time of writing, while a comparable GPT-5.5 agentic run requires additional integration we leave to follow-up work.

\subsection{Agentic Reranking with Autonomous Tool Use}
\label{sec:ablation-agentic}

We now ask a complementary question.
Holding the model, candidate set, and evaluation protocol fixed, does wrapping the same model in an agentic harness change ranking quality?
The harness adds a five-step web + arXiv research routine and autonomous tool access. The routine drives those tool calls, with each step naming a specific tool (arXiv lookup, web search), so the routine and tool access are inseparable by design and single-shot LLM rerank (which has neither component) is the natural control.

On the \nnname~+~BM25 backbone and the original $N{=}200$ subset, we run two settings for each of Opus 4.7 and DeepSeek v4 pro.
\textbf{LLM rerank} sends the original rerank prompt through the provider's API, following the same single-shot rerank protocol used for the LLM rows of Table~\ref{tab:model-sweep}.
\textbf{Agentic rerank} runs the same model through our in-house agent harness, prepending a five-step routine (search arXiv for related papers, use web search to fill in background, disambiguate candidate descriptions when needed, reason step by step over fit, then output the ranking).
Candidate presentation, ranking rules, and output format are identical across the two settings.

\begin{table}[t]
  \centering
  \caption{Agentic vs.\ LLM reranking on the same model.}
  \label{tab:agentic-vs-api}
    \Description{MAP and MRR for two models, Claude Opus 4.7 and DeepSeek v4 pro, each in single-shot LLM rerank mode and in agentic mode with tool access, above a shared no-rerank baseline. The agentic setting is higher for both models.}
  \footnotesize
    \begin{tabular*}{\columnwidth}{@{\extracolsep{\fill}}lcc@{}}
      \toprule
      \textbf{Setup} & \MAP & \MRR \\
      \midrule
      no rerank                                  & 0.260 & 0.302 \\
      \midrule
      \quad+~Opus 4.7 (LLM rerank)               & 0.317 & 0.367 \\
      \quad+~Opus 4.7 (agentic)                  & \textbf{0.323} & \textbf{0.388} \\
      \midrule
      \quad+~DeepSeek v4 pro (LLM rerank)        & 0.310 & 0.366 \\
      \quad+~DeepSeek v4 pro (agentic)           & \textbf{0.313} & \textbf{0.374} \\
      \bottomrule
    \end{tabular*}
  \par
  \parbox{\columnwidth}{\footnotesize\raggedright
    Better result per model group in \textbf{bold}. $\RK{10}/\RK{20}/\RK{100} =
    0.483/0.609/0.829$ in all rows; top-10 reordering preserves the retrieval set.}
\end{table}

Table~\ref{tab:agentic-vs-api} reports the comparison.
For both LLMs the agentic harness delivers a directional \MAP/\MRR gain over the same model in single-shot mode, and the directional consistency across two vendors is the central observation.
We do not yet report paired-bootstrap CIs for the within-model gain (Section~\ref{sec:limitations}).
We also do not compare these agentic-mode numbers cross-model against the LLM-rerank-mode numbers of other models, since that would conflate the harness effect with model-capability differences.

The agentic gain is uneven between the two LLMs.
Opus invokes tools on only about $41\%$ of queries and gains more; DeepSeek invokes them on essentially every query and gains less, suggesting the gain is more about \emph{when} to call tools than raw call count.
Cost, however, follows call count: multiple tool round trips and accumulated reasoning tokens make agentic rerank roughly $5$ to $10$ times more expensive per query than same-model LLM rerank, with wall time moving from seconds to minutes.


\section{Limitations}
\label{sec:limitations}

This paper's core methodological assumption is that the datasets actually cited by peer-reviewed publications serve as a credible silver-label signal for (query, dataset) relevance.
This assumption is not without cost, and we state its boundaries here.

\paragraph{Citation does not equal complete relevance.}
The datasets a paper cites are those the work \emph{used}, not all that are \emph{applicable}; alternative products on the same topic (e.g.\ TRMM, IMERG, and CMORPH for GPM precipitation) are typically not cited and would count as false positives.
We bound this bias by restricting the benchmark to the top 10{,}636 most-cited publications, whose dataset citations more reliably reflect real use, and by reporting LLM Judge $\PK{K}$ as a diagnostic complement to citation recall (Section~\ref{sec:metrics}).

\paragraph{Popularity bias.}
Citation-grounded evaluation rewards datasets with high visibility (MODIS, Landsat, GPM IMERG) and penalises less-used products on the same topic.
The benchmark numbers should therefore be read as an upper-bound estimate of ranking quality on mainstream data products, not as a direct proxy for user-experience quality on long-tail or niche queries.

\paragraph{Tool-augmented agent and potential label leakage.}
Allowing the agent to call web and arXiv tools risks letting it locate the source paper of a query and recover the ground-truth dataset list from that paper's references.
Queries are LLM-rewritten from abstracts into ``I want to\ldots{}'' form rather than echoing title keywords, but instrument/region/temporal-window entities can still seed reverse retrieval, so we cannot fully rule the risk out.
Tighter controls are left to future work.

\paragraph{Evaluation scope and validity.}
The queries and the diagnostic judge are both LLM-produced (\texttt{GPT-5.4} and \texttt{Qwen3.6-35B-A3B}); we do not include a human study of query naturalness or judge--expert agreement.
The agentic experiments evaluate only the web + arXiv tool set, and budget kept them on the original $N{=}200$ subset while the single-shot sweep moved to $N{=}600$, so their within-model gains carry no paired-bootstrap CIs.
Extending to the full test set, the NASA-tool inventory, and human-validated query / judge protocols is left to follow-up work.

\section{Conclusion}
\label{sec:conclusion}

This paper frames geoscience dataset discovery as a trustable and verifiable agentic search task.
Starting from the NASA Earth Observation Knowledge Graph, we build \sysname, a benchmark of 47k citation-grounded pairs.
On top of \sysname we propose a retrieval suite centred on \nnname neural-score correction and a fine-tuned sentence transformer, fused with BM25 through convex combination. We further compare single-shot LLM rerank against an agentic harness that prepends a five-step web + arXiv research routine with autonomous tool calls.
The experiments show two consistent trends.
First, the retrieval suite improves both R@10 and \MRR by more than $5\times$ over the unadapted cosine baseline.
Second, single-shot high-capacity LLM rerank stably improves ranking across two retrieval backbones and all four models in that class, and on the same LLM the agentic harness yields a directional \MAP/\MRR gain over single-shot LLM rerank for both Opus 4.7 and DeepSeek v4 pro on the $N{=}200$ stratified subset. The harness routine is the natural-language protocol that drives the tool calls, so the two are coupled by design and single-shot LLM rerank is the natural control.
One immediate next direction is cost-aware tool-call routing. Opus calls tools on only $41\%$ of queries while DeepSeek does so on $99.5\%$, yet Opus gains more, suggesting that \emph{when} to call tools is a policy question worth studying on its own.


\appendix
\section{Squared-Loss Derivation for the Fusion Weight}
\label{app:alpha-derivation}

We motivate the relative-performance rule in Equation~\eqref{eq:alpha} by asking which convex mixture minimises squared error once both component scorers are fixed, so that $\alpha$ is the only free parameter.

For a training pair $i$ with relevance target $y_i \in [0,1]$, write the min--max normalised neural and lexical scores as $s_{n,i}$ and $s_{\ell,i}$, the fused score as $s_{\alpha,i}=\alpha s_{n,i}+(1-\alpha)s_{\ell,i}$, and the standalone residuals as $e_{n,i}=s_{n,i}-y_i$ and $e_{\ell,i}=s_{\ell,i}-y_i$.
Collecting the standalone mean squared errors $M_n=\frac{1}{N}\sum_i e_{n,i}^2$ and $M_{\ell}=\frac{1}{N}\sum_i e_{\ell,i}^2$ and the residual cross-moment $C=\frac{1}{N}\sum_i e_{n,i}e_{\ell,i}$, the squared-error surrogate is a quadratic in $\alpha$:
\begin{equation}
  \mathcal{L}(\alpha)
  = \frac{1}{N}\sum_{i=1}^{N}\bigl(\alpha e_{n,i}+(1-\alpha)e_{\ell,i}\bigr)^2
  = \alpha^2 M_n + (1-\alpha)^2 M_{\ell} + 2\alpha(1-\alpha)C .
\end{equation}
It is convex, since $\partial^2\mathcal{L}/\partial\alpha^2 = 2(M_n+M_{\ell}-2C) = \frac{2}{N}\sum_i (e_{n,i}-e_{\ell,i})^2 \ge 0$, so whenever the two residual sequences differ the unconstrained minimiser is
\begin{equation}
  \alpha^{\star} = \frac{M_{\ell}-C}{M_n+M_{\ell}-2C} .
  \label{eq:alpha-mse-general}
\end{equation}
Under the common ensemble assumption that the two scorers make approximately uncorrelated residual errors on the training distribution ($C \approx 0$), and treating standalone retrieval ability as an inverse-error quantity, $\pi_n \propto 1/M_n$ and $\pi_{\ell} \propto 1/M_{\ell}$, this reduces to
\begin{equation}
  \alpha^{\star} \approx \frac{M_{\ell}}{M_n+M_{\ell}} = \frac{\pi_n}{\pi_n+\pi_{\ell}} ,
  \label{eq:alpha-mse-uncorrelated}
\end{equation}
which is the relative-performance form of Equation~\eqref{eq:alpha}: the neural scorer receives more weight exactly when the lexical scorer has the larger error.
In the experiments $\pi_n$ and $\pi_{\ell}$ are taken from training-split retrieval metrics rather than squared error, since the final task is ranking rather than calibrated regression; the calculation explains the direction of the rule, not a claim of optimality.


\section{LLM Reranking Prompts}
\label{app:rerank-prompt}

This appendix lists the prompt templates used in the LLM reranking (Section~\ref{sec:ablation-rerank}) and agentic reranking (Section~\ref{sec:ablation-agentic}) experiments.
Placeholders \texttt{\{query text\}}, \texttt{\{title\}}, and \texttt{\{summary\}} are filled from the retrieval stage output.

\subsection*{LLM Rerank}

\begin{promptbox}
\scriptsize
\textbf{System:}

You are an expert Earth science dataset retrieval reranker.

Given a user query and numbered candidate datasets, rank them by relevance.

Rules:
\begin{itemize}
  \item Only return candidate ids from the given candidates.
        Do not invent new ids.
  \item Return the top $K$ most relevant candidate ids only.
  \item Output a single JSON object of the form
        \texttt{\{"ranked": [3, 1, 5, \ldots]\}} listing the ids from most
        to least relevant.
\end{itemize}

\texttt{<query>}

\textit{\{query text\}}

\texttt{</query>}

\texttt{<candidates>}

\texttt{<candidate id="1">}\textit{\{title\} \{summary\}}\texttt{</candidate>}

\texttt{<candidate id="2">}\textit{\{title\} \{summary\}}\texttt{</candidate>}

\texttt{...}

\texttt{</candidates>}
\end{promptbox}

\subsection*{Agentic Rerank}

Identical to the prompt above, with the following routine inserted between the
rules and the \texttt{<query>} block.

\begin{promptbox}
\scriptsize
Before answering, follow this routine:
\begin{enumerate}
  \item Search for papers relevant to the query topic on arXiv.
  \item Use web search to find additional context not covered by step 1.
  \item For candidates whose descriptions are ambiguous or whose relevance
        is unclear, use web search to clarify what each candidate measures.
  \item Think step by step: which candidates best match the query's specific
        requirements.
  \item Output the ranked JSON.
\end{enumerate}
\end{promptbox}

\begin{acks}
This work was supported in part by the \grantsponsor{NSF}{U.S. National Science
Foundation}{https://www.nsf.gov/} under grants \grantnum{NSF}{CAIG RISE-2615702},
\grantnum{NSF}{IIS-2520978}, \grantnum{NSF}{CAIG RISE-2530596},
\grantnum{NSF}{CAIG RISE-2615703}, \grantnum{NSF}{CAIG RISE-2530597}, and
\grantnum{NSF}{CAREER AGS-2541302}.
We thank \href{https://modal.com}{Modal Labs} for providing compute credits.
\end{acks}

\section*{GenAI Usage Disclosure}

In compliance with the ACM Policy on the use of generative AI tools, we disclose the following.
\textbf{Benchmark construction:} Each query in \sysname is generated by \texttt{GPT-5.4} (temperature 0.2) from a publication abstract; ground-truth labels are sourced independently from NASA EO-KG \texttt{USES\_DATASET} edges and are \emph{not} LLM-produced (prompt in the \sysname dataset card, \url{https://huggingface.co/datasets/HamiltonMYu/NASA-EO-Bench}).
\textbf{Evaluation:} The LLM-as-a-Judge $\PK{K}$ metric (Section~\ref{sec:metrics}) uses Qwen3.6-35B-A3B via vLLM.
\textbf{Reranking:} LLM rerank and agentic rerank experiments call OpenAI o4-mini, GPT-5.4, and GPT-5.5, Claude Opus 4.7, and DeepSeek v4 pro via hosted APIs or an in-house agentic harness (prompts in Appendix~\ref{app:rerank-prompt}).
\textbf{Code and writing:} Anthropic \texttt{Claude} assisted with experiment scripts, \LaTeX{} formatting, and manuscript editing; the authors verified every claim, number, and result reported here.


\bibliographystyle{ACM-Reference-Format}
\balance
\bibliography{refs}

\end{document}